# Sustainable Load Balancing for Wireless Networks With Renewable Energy Sources

Mustafa Mohammed Hasan Alkalsh
*Poznan University of Technology*
Poznan, Poland
*Nokia Solutions and Networks*
Wroclaw, Poland
ORCID: 0000-0001-8777-9130

Adam Samorzewski
*Poznan University of Technology*
Poznan, Poland
ORCID: 0000-0003-4138-5544

Adrian Kliks
*Poznan University of Technology*
Poznan, Poland
ORCID: 0000-0001-6766-7836

*Abstract*—Future wireless networks powered by renewable energy sources and storage systems (e.g., batteries) require energy-aware mechanisms to ensure stability in critical and high-demand scenarios. These include large-scale user gatherings, especially during evening hours when solar generation is unavailable, and days with poor wind conditions that limit the effectiveness of wind-based energy harvesting. Maintaining network performance under such constraints, while preserving stored energy, remains a key challenge. This work proposes an enhanced Proactive-Reactive Load Balancing algorithm that integrates energy conditions into mobility management. By leveraging standardized mobility events, the algorithm optimizes traffic distribution and energy utilization (avoiding complete drainage of stored energy), thereby preventing service degradation. Simulations show improved energy sustainability and network performance under congestion and limited solar availability.

*Index Terms*—Sustainable wireless networks, energy-aware mobility management, load balancing, renewable energy sources

## I. Introduction

The rapid densification of 5G and beyond networks, driven by ultra-low latency and high throughput demands, has significantly increased energy consumption across Radio Access Networks (RANs). This energy surge has motivated widespread efforts to design energy-efficient mechanisms, especially for deployments powered by Renewable Energy Sources (RES) and storage systems (e.g., batteries).

Recent studies have addressed cells' energy savings through advanced sleep scheduling [1], [2], and energy-aware resource orchestration that optimizes cross-domain resource allocation in heterogeneous environments, including satellite and aerial layers [3]. However, most existing work centers on static configurations or user equipment (UE)-side energy optimization, leaving mobility management underexplored from a cell energy sustainability perspective. Meanwhile, off-grid and hybrid-powered base stations face volatile energy availability. Events with dense user activity can deplete RES cell batteries quickly during low solar or wind periods, a challenge that conventional mobility mechanisms fail to address. While carbon footprint reduction is well-discussed [4], dynamic energy-aware mobility remains underexplored.

The aim of this work is to analyze how real-time energy sustainability indicators can be incorporated into mobility management decisions for small cells powered by RES. Therefore, this paper proposes an enhanced Proactive-Reactive Load Balancing (PRLB) algorithm that embeds energy-awareness into mobility management. Specifically, introduce a continuous energy sustainability index—derived from real-time battery state of charge, net energy balance, and load congestion—to scale Cell Individual Offset (CIO) values, influencing A3-based handovers [5]. Additionally, a two-level power-saving mechanism is activated when RES cells operate under constrained energy, allowing for proactive traffic shedding to extend operational longevity. However, this work focuses on energy-aware power-saving mechanisms for RES-powered cells under low-battery conditions, while future work will explore proactive mobility strategies that leverage periods of high renewable energy availability.

Unlike conventional load-balancing techniques that react solely to traffic conditions, the proposed approach jointly considers energy and load dynamics to enable smart, sustainable handover control in dense, off-grid small-cell networks. In essence, the novelty of the paper lies in its integration of real-time energy sustainability metrics into mobility management, enabling dynamic, energy-conscious control over UE association and CIO adjustment. Simulation results show improved traffic distribution and energy preservation without compromising service availability.

The paper is organized as follows: Section II introduces the system model; Section III details the proposed method;

This paper has been accepted for publication in the IEEE International Conference on Pervasive Computing and Communications (PerCom 2026). The final version will be available via IEEE Xplore.

This work was funded by the MEiN "Applied Doctorate" Program (Agreement No. DWD/6/0057/2022) of the Ministry of Science and Higher Education, Republic of Poland. It was also supported by the National Science Centre (NCN) in Poland under project no. 2021/43/B/ST7/01365 within the OPUS program.

Section IV presents evaluations; and Section V concludes.

## II. System Model

The considered system model encompasses network topology with renewable energy integration, including the dense urban layout described below, together with radio propagation, mobility, and traffic models, along with the baseline PRLB scheme for evaluating the proposed enhancement. The scenario, as shown in Fig. 1, considers a dense urban layout with 17 small cells $C$ operating on inter-frequency NR bands between $3.4\,\text{GHz}$ to $3.6\,\text{GHz}$. Each site is sectorized into three $120°$ sectors with directional antennas. A focused cluster of 9 cells (central region) is analyzed, including several base stations powered exclusively by RES, such as solar or wind, and marked with a battery icon. The map is built from real-world urban street data using the OSMnx library [6]. UEs follow target-based waypoint mobility with mixed-speed profiles. RES cells, $i \in C$, harvest and store energy via batteries with SoC, $s_i \in [0,1]$, is updated at each time step according to:

$$s_i(t) = \frac{E_{\text{batt},i}(t-1) + E_{\text{g},i}(t) - E_{\text{c},i}(t)}{E_{\text{batt},i}^{\max}}, \quad (1)$$

where $E_{\text{batt},i}$ is the energy stored in the battery. Next, $E_{\text{g},i}$ and $E_{\text{c},i}$ are the energies harvested by RES generators, and consumed for transmission and control overhead, respectively. Finally, $E_{\text{batt},i}^{\max}$ is the maximum energy capacity for a single battery. The system assumes off-grid operation (no power fallback) and constant transmission power. The mathematical models used, which describe energy consumption and generation processes, are based on a set of practical assumptions from [7].

However, the Radio propagation follows the UMi Street Canyon model (3GPP TR 38.901, Table 7.4.1-1) [8], including path loss, shadowing, and fading. UEs in RRC-connected mode measure reference signal received power (RSRP) and Signal-to-interference-plus-noise ratio (SINR), triggering standard 3GPP handover events (A1, A2, A3) [9]. Events A1–A3 manage UE handovers: A1 triggers when the serving signal exceeds a threshold (TTT), A2 when it drops below, and A3 when a neighbor's signal plus offset surpasses the serving one. CIO offsets dynamically adjust cell-edge behavior. The CIO dynamically adjusts handover aggressiveness.

UE attachment requests follow a Poisson process with exponentially distributed inter-arrival times. Traffic demands are heterogeneous, and resource scheduling is performed based on the Channel Quality Indicator (CQI). Cells monitor both instantaneous and average Physical Resource Block (PRB) utilization, which feeds into the load balancing algorithm. The baseline PRLB algorithm [5] operates on two time scales:

- **Proactive:** Triggered when PRB usage exceeds a critical threshold (e.g., 98%), CIOs are adjusted to balance load.
- **Reactive:** average load is checked periodically (e.g., every 4 minutes); if over 65%, initiate load redistribution by adjusting CIOs.

However, PRLB ignores energy availability, which may lead to excessive RES battery drain during congestion or low generation, which is an issue addressed by the proposed enhancement in Section III.

## III. Proposed Energy-Aware PRLB Enhancement

The energy-aware enhancement introduces a dynamic weighting mechanism in the proposed ePRLB to account for the energy conditions of RES cells. For each RES cell $i$, the algorithm computes a continuous sustainability score $E_i \in [-1, 1]$ based on its SoC, $s_i \in [0, 1]$, net power flow, and mid-range discharge behavior. However, ePRLB classifies RES cells based on its roles: As a neighbor ($n$Cell) considered for offloading by overloaded cells. Or, as a serving cell entering power-saving mode due to critical SoC and energy deficit. For neighbor cells, the PRLB-generated CIO bias used during A3-triggered handover evaluation is energy-aware scaled according to $E_i$. This signed scaling attenuates, neutralizes, or inverts the PRLB bias, thereby reducing the effective visibility of energy-constrained RES cells while reinforcing handovers toward cells with favorable energy conditions. This energy-aware scaling incorporates several key features, including SoC sensitivity, net energy trends, and mid–range discharge behavior.

*(a) Logistic SoC Normalization:* The contribution of SoC to the energy sustainability score $E_i$ is modeled as:

$$B(s_i) = 2\sigma\left(k_{\text{soc}} \frac{s_i - m}{r}\right) - 1, \quad \sigma(x) = \frac{1}{1 + e^{-x}}, \quad (2)$$

where $m = (s_H + s_L)/2$ and $r = s_H - s_L$. The shaping factor $k_{\text{soc}}$ controls the slope of the transition. The thresholds are $s_L = 0.2$ (critical) and $s_H = 0.4$ (sufficient) are adopted based on SoC management principles discussed in [10], [11], where maintaining battery health and avoiding deep discharge are emphasized. The proposed approach ensures that $B(s_i) \approx -1$ when $s_i < s_L$ discouraging handovers); then $B(s_i) \approx +1$ when $s_i > s_H$ favoring handovers; and finally, $B(s_i) \in (-1, +1)$ in between, offering smooth adaptation.

*(b) Net–Power Adaptation:* To reflect short-term energy dynamics at the RES cell, specifically, the balance between instantaneous energy generation and consumption, the following normalized power adaptation function is used:

$$A(E_{g,i}, E_{c,i}) = \beta \tanh\left(\frac{E_{g,i} - E_{c,i}}{\xi}\right), \quad (3)$$

where $\xi$ is the maximum absolute net power difference in the input vector, clipped to a small minimum (e.g., $10^{-6}$) to avoid division by zero. The coefficient $\beta \in [0,1]$ adjusts the influence of net charging/discharging. The bounded tanh ensures numerical stability, as commonly used in hybrid RES control systems [12].

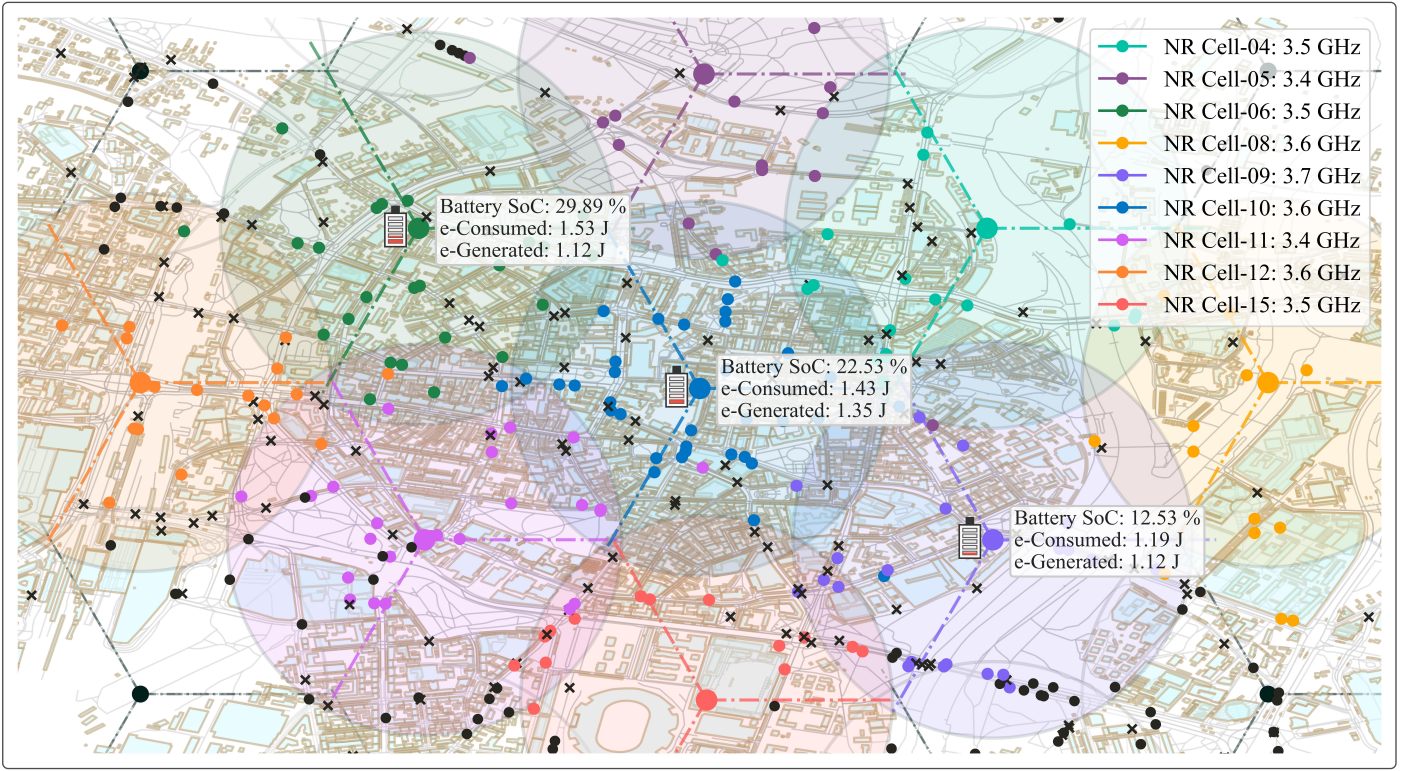

Fig. 1. Simulation scenario in Poznań, Poland, with active load-balancing, showing RES cells with battery SoC, energy consumption, and generation

*(c) Mid–Range Discharge Attenuation:* To prevent overreaction in moderate SoC conditions, where neither full offloading nor complete inactivity is appropriate, a Gaussian attenuation term is introduced:

$$Z(s_i) = 1 - \lambda \exp\left[-\frac{(s_i - \mu)^2}{2\sigma^2}\right], \quad (4)$$

with $(\mu, \sigma) = (0.35, 0.12)$ empirically selected to center the attenuation effect around mid-range SoC levels. The parameter $\lambda \in [0, 1]$ controls the attenuation strength. This mechanism moderates energy response to avoid excessive CIO scaling fluctuations and is inspired by fuzzy energy control principles for graceful performance degradation [13], [14].

*(d) Composite Energy Index:* The final energy state index is:

$$E_i = \Psi\left([B(s_i) + A(E_{g,i}, E_{c,i})] \cdot Z(s_i)\right), \quad (5)$$

with:

$$\Psi(x) = \begin{cases} -1, & x < -1, \\ x, & |x| \leq 1, \\ +1, & x > 1. \end{cases} \quad (6)$$

This continuous, bounded index modulates CIO values during ePRLB operations to bias UE handovers away from low-energy cells, as shown in Algorithm 1. However, this mechanism supports context-aware load balancing consistent with energy-adaptive networks [15], [16].

**Algorithm 1** Energy-Aware CIO Scaling for RES Cells
**for all** $j \in \mathcal{C}$ **do**
  **if** $j$ is RES **then**
    $B(s_j) \leftarrow 2\sigma(k_{\text{soc}}(s_j - m)/r) - 1$
    $A \leftarrow \beta \tanh((E_{g,j} - E_{c,j})/\xi$
    $Z \leftarrow 1 - \lambda \exp(-(s_j - \mu)^2/2\sigma^2)$
    $E_j \leftarrow \Psi((B + A) \cdot Z)$
    $\text{CIO}_j \leftarrow \text{CIO}_j \cdot E_j$
  **end if**
**end for**

### A. Dual-Phase Triggering for Energy Preservation

In addition to influencing offloading decisions, RES cells with $s_i \leq s_H$ and a net energy deficit $E_{c,i} > E_{g,i}$, may be scheduled for power-saving operations. The proposed mechanism adopts a dual-phase strategy that dynamically triggers power-saving actions based on SoC severity and recent energy trends:

- **Reactive Phase:**
  – If $s_i \in [m, s_H]$, and the condition $s_i \leq s_H$ and $E_{c,i} > E_{g,i}$ has been sustained over the last $T$ interval, the cell is scheduled for a power-saving check every $T$.
  – If $s_i \leq m$, and the condition remains true over the last $T/2$, the cell is scheduled for a check every $T/2$.

- If the monitored RES cell no longer satisfies the condition within the corresponding interval ($T$ or $T/2$), it is removed from the power-saving schedule.

- **Proactive Phase:** If $s_i \leq s_L$, the power-saving mechanism is triggered immediately, bypassing the monitoring interval.

### B. Energy Reduction Ratio Estimation

The ePRLB computes the required consumption energy reduction ratio $\gamma_i$ to determine how much energy a RES cell should aim to conserve under critical conditions. This estimation involves the following components:

1) **Deficit Ratio:** Represents the severity of energy shortfall based on generation and consumption:

$$\delta_i = \max\left(0, 1 - \frac{E_{g,i}}{E_{c,i}}\right).$$

2) **SoC-Based Cap:** Imposes an upper limit on allowable energy consumed reduction based on the State of Charge (SoC). Lower SoC levels permit more aggressive energy savings:

$$\mathrm{cap} = \begin{cases} 0.75, & s_i \leq 10\% \\ 0.55, & 10\% < s_i \leq 20\% \\ 0.35, & 20\% < s_i \leq 30\% \\ 0.20, & 30\% < s_i \leq 40\% \\ 0.00, & \text{otherwise} \end{cases}$$

3) **Congestion Adjustment:** If more than 60% of neighbors are overloaded (PRB usage > 85%), the cap is reduced by 25%.
4) **Final Ratio:** $\gamma_i = \min(\delta_i, \mathrm{cap})$.

However, it is worth noting that the cap values are empirically chosen to reflect increasing urgency as SoC declines, and can be adjusted based on deployment-specific energy patterns. Additionally, future work may explore AI-driven methods to adapt these thresholds dynamically in response to historical or contextual energy data.

### C. Load Reduction Execution

Given energy reduction ratio $\gamma_i$, the energy to reduce is $\Delta E_i = \gamma_i \cdot E_{c,i}$. The RES cell offloads UEs based on energy contribution and distance until $\Delta E_i$ is met. CIOs are updated using the PRLB logic based on the last RSRP measurements [5].

**Algorithm 2** Power-Saving for RES Cells

**for** $i \in \mathcal{C}$ **do** {At each time step $t = 1$s}
  **if** $m < s_i \leq s_H$ **and** $E_{c,i} > E_{g,i}$ **then**
    Add or retain $i$ in `IntervalList`
  **else if** $s_i \leq m$ **and** $E_{c,i} > E_{g,i}$ **then**
    Add or retain $i$ in `HalfIntervalList`
  **else**
    Remove $i$ form List **if** $i$ in any
  **end if**
  **if** $s_i \leq s_L$ **or** ($i$ in `IntervalList` **and** $t \bmod T = 0$) **or** ($i$ in `HalfIntervalList` **and** $t \bmod (T/2) = 0$) **then**
    $\gamma_i \leftarrow$ ESTIMATEENERGYREDUCTION($s_i, E_{g,i}, E_{c,i}$)
    EXECUTEPOWERSAVING($\gamma_i$)
  **end if**
**end for**
**function** ESTIMATEENERGYREDUCTION($s_i, E_{g,i}, E_{c,i}$)
  $\delta_i \leftarrow \max(0, 1 - \frac{E_{g,i}}{E_{c,i}})$; cap $\leftarrow$ SoC-based
  **if** over 60% neighbors congested **then** cap $\leftarrow$ cap$\cdot 0.75$
  **end if**
  **return** $\min(\delta_i, \mathrm{cap})$
**end function**
**function** EXECUTEPOWERSAVING($\gamma_i$)
  $\Delta E_i = \gamma_i \cdot E_{c,i}$
  Check UEs until $\Delta E_i$ reached
  Update CIO via PRLB
**end function**

## IV. RESULTS DISCUSSION

To evaluate the proposed approach, three identical simulation environments based on the city center of Poznań were executed over 19 minutes, repeated more than 20 times with varying user congestion and renewable energy availability. The first environment was without load balancing (Without LB), the second employed the PRLB algorithm, while the third one used the proposed ePRLB. The results in Fig. 2 show that PRLB—while effective in balancing traffic—reduces RES cell SoC by approximately 10.8% compared to Without-LB. In contrast, ePRLB preserves energy more efficiently, improving average SoC by 3.15% over Without-LB baseline. This indicates that load balancing algorithms such as PRLB may inadvertently drain energy-limited cells, while ePRLB ensures more sustainable energy usage without compromising load distribution. While the average SoC is used as the primary indicator in this study, it may mask variability across individual RES-powered cells. A detailed analysis of SoC dispersion and fairness among cells is part of ongoing work.

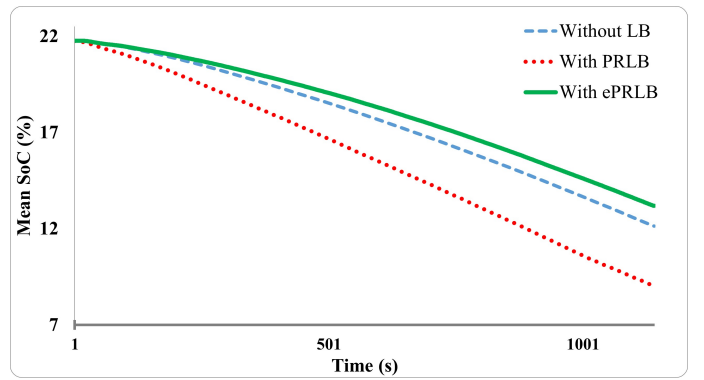

Fig. 2. Mean SoC of RES cells over the simulation time.

Fig. 3 further highlights how ePRLB dynamically shifts traffic load, quantified as an aggregated PRB utilization index of the analyzed cells, away from critically charged RES cells, achieving a more balanced load more intelligently than PRLB. However, while results are based on 19-minute traces, they underline the importance of energy-aware mobility management—particularly under high density of users and variable energy generation scenarios—with expected stronger effects over longer durations.

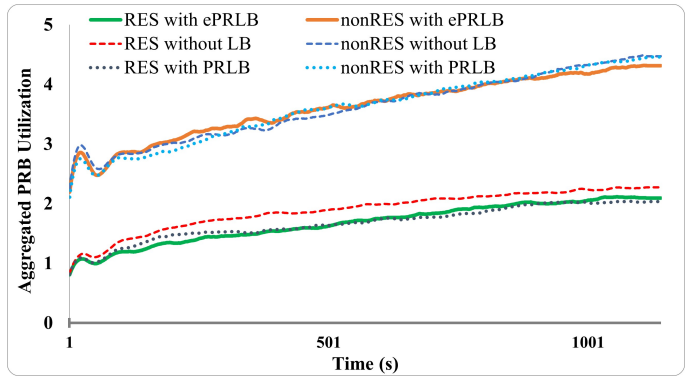

Fig. 3. Distribution of user traffic across RES and non-RES cells during the simulation periodm, where traffic is quantified as an aggregated PRB utilization index computed across the analyzed cells.

## V. Conclusions

This paper emphasizes the need for energy-aware mechanisms in dense networks powered by renewable energy sources. While conventional load balancing improves traffic distribution, it may inadvertently accelerate battery depletion in RES cells. The proposed ePRLB incorporates real-time energy indicators into mobility management, leading to smarter and more sustainable handover decisions. Simulation results show that ePRLB preserves more battery energy compared to PRLB, even within short operational windows—demonstrating its effectiveness in maintaining energy resilience under dynamic network and generation conditions. Mobility management metrics such as ratios of handover ping-pong, handover failure, radio link failure, attach failure, and overload duration were extensively analyzed in the original PRLB framework. In this work, the focus is placed on introducing energy-awareness into mobility control via ePRLB; a comprehensive evaluation of these mobility metrics under the proposed scheme is part of ongoing work. As future work, AI/ML techniques will be explored to dynamically adapt energy control policies based on historical and contextual data.

## References

[1] D. Renga, Z. Umar, and M. Meo, "Trading off delay and energy saving through advanced sleep modes in 5g rans," *IEEE Transactions on Wireless Communications*, vol. 22, no. 11, pp. 7172–7184, 2023.

[2] L. Dorval, C. Gueguen, and G. Terrier, "Energy-aware meta sleep scheduler, an alternative to drx in 5g networks," *IEEE Access*, vol. 13, pp. 1–14, 2025.

[3] P. Zhang, Z. Li, M. Guizani, N. Kumar, K. Yu, and J. Wang, "Energy aware space-air-ground integrated network resource orchestration algorithm," *IEEE Transactions on Vehicular Technology*, vol. 73, no. 12, pp. 18 950–18 960, 2024.

[4] F. Antão, H.-R. Chi, and D. Corujo, "Toward decarbonized network function virtualization-based ict management," *IEEE Transactions on Green Communications and Networking*, vol. 7, no. 2, pp. 523–537, 2025.

[5] M. M. H. Alkalsh and A. Kliks, "Scalable mobility proactive-reactive load balancing (prlb) algorithm for high-density environments of future wireless networks," *IEEE Access*, 2025.

[6] G. Boeing, "Modeling and analyzing urban networks and amenities with osmnx," *Geographical Analysis*, 2025.

[7] A. Samorzewski, M. Deruyck, and A. Kliks, "Energy consumption in res-aware 5g networks," in *GLOBECOM 2023-2023 IEEE Global Communications Conference*. IEEE, 2023, pp. 1024–1029.

[8] 3GPP, TR-38.901, "Study on channel model for frequencies from 0.5 to 100 GHz (Release 17)," 3rd Generation Partnership Project (3GPP), Tech. Rep. V17.0.0, 2022, 3GPP Technical Report (TR).

[9] 3GPP, TS-38.331, "NR; Radio Resource Control (RRC); Protocol Specification (Release 17)," 3rd Generation Partnership Project (3GPP), Tech. Rep. V17.4.0, 2023, 3GPP Technical Specification (TS).

[10] M. A. Hannan, M. S. H. Lipu, A. Hussain, and A. Mohamed, "A review of lithium-ion battery state of charge estimation and management system in electric vehicle applications: Challenges and recommendations," *Renewable and Sustainable Energy Reviews*, vol. 78, pp. 834–854, 2018.

[11] G. L. Plett, *Battery Management Systems, Volume II: Equivalent-Circuit Methods*. Norwood, MA: Artech House, 2015.

[12] C. Zhang, J. Jiang, W. Zhang, and P. Liu, "Power and energy management strategies for hybrid energy storage system in electric vehicles—a review," *Journal of Power Sources*, vol. 448, p. 227418, 2020.

[13] D. Wu, L. Wang, X. Zhang, and H. Peng, "Energy management strategy based on gaussian membership functions for plug-in hybrid electric vehicles," *IEEE Transactions on Vehicular Technology*, vol. 69, no. 6, pp. 6023–6035, 2020.

[14] M. Alipour, Q. Shafiee, and J. M. Guerrero, "Distributed control of microgrids based on gaussian and fuzzy membership functions," *IEEE Transactions on Smart Grid*, vol. 9, no. 2, pp. 1263–1272, 2018.

[15] J. Wu, S. Dai, and H. Wang, "Energy-aware load balancing for wireless sensor networks based on adaptive fuzzy logic," *Sensors*, vol. 20, no. 3, p. 822, 2020.

[16] Y. Chen, T. Liu, and D. W. Gao, "Unified state-of-energy index for battery–supercapacitor hybrid systems," *IEEE Transactions on Industrial Electronics*, vol. 68, no. 6, pp. 5062–5072, 2021.